\documentclass[12pt]{article}
\usepackage{epsfig,psfig,amssymb,cite}

\newcommand{\rv}{{\bf r}}
\newcommand{\rvp}{{\bf r}_\parallel}

\newcommand{\ri}{\right)}
\newcommand{\li}{\left(}

\begin{document}
\setlength{\unitlength}{0.2cm}

\title{Universal shape ratios for polymers grafted at a flat
surface}

\author{
  \\
  {\small Maria Serena Causo}             \\[-0.2cm]
  {\small\it INFM-NEST and Scuola Normale Superiore}\\[-0.2cm]
  {\small\it Piazza dei Cavalieri, 7}  \\[-0.2cm]
  {\small\it I-56100 Pisa, ITALY}          \\[-0.2cm]
  {\small Internet: {\tt causo@sns.it}}     \\[-0.2cm]
  {\protect\makebox[5in]{\quad}}  
  \\
}
\vspace{0.5cm}

\maketitle
\thispagestyle{empty}   

\def\spose#1{\hbox to 0pt{#1\hss}}
\def\ltapprox{\mathrel{\spose{\lower 3pt\hbox{$\mathchar"218$}}
 \raise 2.0pt\hbox{$\mathchar"13C$}}}
\def\gtapprox{\mathrel{\spose{\lower 3pt\hbox{$\mathchar"218$}}
 \raise 2.0pt\hbox{$\mathchar"13E$}}}

\vspace{0.2cm}

\begin{abstract}
We consider dilute non-adsorbed polymers grafted at an 
impenetrable surface and compute several quantities which
characterize the polymer shape: the asphericity and the ratios
of the eigenvalues of the radius-of-gyration tensor.
The results are only slightly different from those obtained
for polymers in the bulk, showing that the surface has
little influence on the polymer shape.
\end{abstract}
\clearpage

\newcommand{\be}{\begin{equation}}
\newcommand{\ee}{\end{equation}}
\newcommand{\bea}{\begin{eqnarray}}
\newcommand{\eea}{\end{eqnarray}}
\newcommand{\<}{\langle}
\renewcommand{\>}{\rangle}

\newcommand{\R}{\hbox{{\rm I}\kern-.2em\hbox{\rm R}}}

\newcommand{\reff}[1]{(\ref{#1})}

\section{Introduction}
It has been known since 1934 that polymers are not istantaneously 
spherical in shape~\cite{Kuhn}.
Since then a lot of work has been devoted to quantify the average
deviation from spherical symmetric shape in the limit of large degree
of polymerization, since this deviation plays an important role
in the hydrodynamic behaviour of dilute polymer solutions~\cite{Kramers}.
There are several observables which quantify the shape 
of polymers. 
One considers the (squared) radius-of-gyration tensor and its
eigenvalues $q_1,\dots,q_d$, which give the square lengths
of the polymer along the principal axes of inertia~\cite{solc}.
The ratios $\< q_i\>/\<q_j\>$ or, alternatively, $\< q_i/q_j\>$ are
universal and may be used to give a universal characterization 
of the polymer shape. Alternatively, one may use the shape asphericity that 
allows to distinguish between spherical and rodlike configurations.
The equivalent ellipsoid has been experimentally determined in
a small-angle x-ray scattering experiment for globular proteins~\cite{exp}.

The shape of Gaussian polymers has been extensively studied.
The shape asphericity, which in its two variants is proportional to the istantaneous
variance of the average of eigenvalues of the radius-of-gyration tensor,
has been exactly computed~\cite{Eisen,Rudnick_86}.
Instead, the ratios of the eigenvalues $q_i$ are not exactly known.
There are perturbative estimates at second order in $1/d$~\cite{Rudnick_ratios},
which are, however, not very predictive in the physically relevant case $d=3$,
and several estimates from numerical simulations~\cite{Rudnick_ratios,Mazur_73,Bishop_89,
Bruns,Sciutto_94,Zifferer_94,Sciutto_95}. 
The numerical estimates are in good agreement with each other.

In the case of polymers with excluded-volume interactions, asphericity
has been computed to second order in $\epsilon = 4-d$ in Ref.~\cite{Kremer_92}
and by numerical simulations in Refs.~\cite{Bishop_89,Bishop_88,
Batoulis_89,Cannon_91,Kremer_92,Sciutto_96,Zifferer_98,Zifferer_99,
Zifferer_01,Timoshenko_02}, 
while the ratios of the eigenvalues of the radius-of-gyration tensor
have been computed by Monte Carlo simulations in Refs.~\cite{Mazur_73,Bruns,
Sciutto_96,Zifferer_98,Zifferer_99,Zifferer_01,Timoshenko_02}.
Here we consider the case of polymers with excluded-volume
interactions grafted at an impenetrable interface at the ordinary
transition~\cite{Diehl_rev,Diehl_97,Eisen_book}.
Polymers are modeled as $N$-step self-avoiding walks
on a simple cubic lattice.
We perform an extensive Monte Carlo simulation using a variant
of the pivot algorithm~\cite{Lal,McDonald,Sokal}, the cut-and-permute
algorithm, which has been shown to be optimal 
for non-adsorbed polymers grafted at a surface~\cite{cp}.
The efficiency of the algorithm allows us to simulate long
walks with $N\le 32000$ and to study how the presence
of the surface changes the shape of the polymer. 

We find that the main effect of the presence of the surface is to
increase the length of the longest axis of the inertia ellipsoid
with respect to the other two, whose relative length remains, instead, of the
same order.  
Nevertheless, being this increment in asphericity only of a few percents,
we expect that it can only be detected in high-resolution experiments.
%
The paper is organized as follows. In Sec. 2 we define the observables
that will be computed.
In Sec. 3 we present our Monte Carlo results, paying due attention
to the corrections to scaling. In Sec. 4 we compare our results 
with those for polymers in the bulk.

\section{Definitions}
We model the polymer as an $N$-step self-avoiding walk on a simple
cubic lattice, each step representing a single monomer.
The polymer has one of the two endpoints grafted at an impenetrable
flat surface of equation $z =0$, so that the accessible space is
${\mathbb R}^3_+ = \{ \rv = (\rvp,z) \, : \, \rvp \in {\mathbb R}^2, z \ge 0\}$.
Let us denote by ${\bf \omega} = \{\omega_0,\dots , \omega_N\}$ a
generic polymer configuration, where $\omega_i = (\omega_i^1,
\omega_i^2,\omega_i^3)\in {\mathbb R}^3_+$ is the position vector of the
$i$-th monomer and $\omega^3_0 = 0$, and
let us call $W_N$ the set of allowed configurations.

One can introduce several invariant quantities which describe
the shape of the polymer.
In generic dimension $d$ we define the radius-of-gyration or shape tensor
\be
Q_{\alpha \, \beta} =\frac{1}{N} \sum _{i>j} (\omega_i^\alpha 
-\omega_j^\alpha) (\omega_i^\beta -\omega_j^\beta)\, ,
\ee
whose trace concides with the squared radius of gyration
\be
R_g^2 = {\rm Tr}\, Q\, .
\ee
If we denote with $\{q_1,q_2,\dots,q_d\}$ the set of 
eigenvalues of $Q$, ordered in such a way that 
$q_1 \le q_2 \le\dots\le q_d$,
and with $\bar q = \frac{1}{d}{\rm Tr}\, Q $ their average,
we can define the shape asphericity of the polymer as
\be
A_d(\omega) = 
\frac{1}{d\,(d-1)}\sum_{\alpha =1}^d \frac{(q_\alpha -\bar q)^2}{\bar q^2} 
= \frac{d}{d-1}\frac{{\rm Tr}\,(\widehat Q_d^2)}{({\rm Tr} \, Q)^2}\,,
\label{asp_1}
\ee
where $\widehat Q = Q -\frac{1}{d} ({\rm Tr} \, Q)\, I_d$ and
$I_d$ is the identity matrix.
Since the eigenvalues of $Q$ correspond to the squared lengths of the equivalent
ellipsoid axes and the eigenvectors represent the axes directions, 
$A_d(\omega) =0$ for a spherically symmetric configuration---in this case all 
eigenvalues take the same value---while $A_d(\omega) =1$ if the 
polymer is completely elongated and all eigenvalues except one vanish.
The mean asphericity is obtained by averaging $A_d(\omega)$ over
the allowed polymer conformations
\be
 A_d^N  =\frac{1}{c_N} \sum_{{\bf \omega} \in W_N} A_d(\omega)\, ,
\ee
where $c_N$ is the number of allowed configurations and behaves 
as $c_N \sim \mu^N N^{\gamma_1 -1}$
in the critical limit $N\to \infty$. Here $\mu$ is the critical fugacity,
which does not depend on the presence of the surface, and $\gamma_1$
is the susceptibility critical surface exponent 
$\gamma_1 = 0.679(2) $ (Ref.~\cite{Grass}), $\gamma_1 \approx 0.680$ (Ref.~\cite{Diehl_98}).
Other invariant quantities are often considered in order
to characterize the polymer shape.
For instance, one may consider
\be
 \widehat A_d^N  =
\frac{d}{d-1}\frac{\<{\rm Tr}\,(\widehat Q_d^2)\>}{\<({\rm Tr} \, Q)^2\>}\,,
\ee
or the relative magnitudes 
of the averages of the eigenvalues, $\<q_2\>/\<q_1\>$, $\<q_3\>/\<q_1\>$.
The advantage of studying these quantities is that they do not require
computing an average of the ratio of highly fluctuating quantities.

In the limit of large degree of polymerization $N\to \infty $,
these quantities reach universal values with corrections of order
$N^{-\Delta}$,
\bea
A_d^N  &=& A_d\, \li 1+\frac{B}{N^\Delta}+\dots \ri\, ,\\
\widehat A_d^N  &=& {\widehat A}_d \, 
\li 1+\frac{\widehat B}{N^\Delta}+\dots \ri\, ,\\
\frac{\<q^N_2\>}{\<q^N_1\>} &=& L^2_{2,1}  
\li 1+\frac{l_{2,1}}{N^\Delta}+\dots \ri\, ,\\
\frac{\<q^N_3\>}{\<q^N_1\>} &=& L^2_{3,1}  \li 1+\frac{l_{3,1}}{N^\Delta}+\dots \ri\, ,
\label{univ}
\eea
where $\Delta$ is the leading correction-to-scaling exponent.
On the other hand, the eigenvalues of the gyration tensor scale as
\be
\<q^N_i\> = b_i N^{2\nu}\li 1 +\frac{c_i}{N^\Delta}+\cdots \ri \qquad i=1,2,3 \, .
\label{eigenvalues}
\ee
In the following section we present a Monte Carlo study of the 
universal behaviour of these quantities.
\section{Monte Carlo study of the shape asphericity}
We computed the eigenvalues of the shape tensor and the asphericities
by Monte Carlo simulations.
The system was modeled as a self-avoiding walk grafted at the origin
of the half cubic lattice ${\mathbb Z}^3_+$.
In order to sample the phase space $W_N$, we used a nonlocal 
Monte Carlo algorithm, the {\em cut-and-permute} algorithm~\cite{cp}.
It is an optimal algorithm for studying 
polymers grafted at an impenetrable surface at the ordinary
transition, since autocorrelation times in CPU units for global observables
are simply proportional to the degree of polymerization $N$.
The algorithm has, therefore, the same optimal behaviour as the
pivot algorithm for polymers in the bulk.

The system in the bulk is known to have strong corrections to scaling~\cite{note}, since
the leading correction-to-scaling exponent is
$\Delta_{\rm bulk} \approx 0.5$.
It is expected that polymers in the presence of a surface show analogous
corrections, and, indeed, our estimates of the leading correction-to-scaling exponent
are close to $0.5$.

Since the presence of large corrections to scaling introduces systematic 
errors in the estimates for finite $N$, we studied the system for large $N$,
$1000 \le N \le 32000$.
At first, we studied the behaviour of the eigenvalues of the
shape tensor and of its trace $R_g^2$.
The results are displayed in Table~\ref{eigen}.
These quantities are expected to increase 
as $N^{2\,\nu}$, being $\nu$ the same critical 
exponent as in the bulk.
In order to check this prediction, we performed fits of the form
\bea
\<q_i\>_N& =& b_i N^{2\, \nu} \qquad i=1,2,3 \, ,\nonumber\\
\<R_g^2\>& =& b N^{2\, \nu} \, , \label{fit1}
\eea
where $ b$, $b_i$, and $ \nu$ are free parameters.
In Eq.~\reff{fit1} corrections to scaling are neglected.
We expect, therefore, a systematic error in our estimates.
In order to take into account the effect of corrections to scaling,
we have performed different fits, considering only data with 
$N\ge N_{\rm min}$. 
The effective exponents which are obtained
decrease with increasing $N_{\rm min}$ for all values of 
$N_{\rm min}$ considered, so that we can only give an upper bound
for the exponent $\nu$.
The estimates which are less affected by corrections to scaling
appear to be those corresponding to the largest eigenvalue.
Therefore, taking as an upper bound for $\nu$ the estimate
from the fit of $\< q_3\>$ with $N_{\rm min} = 16000$, we obtain
$\nu \le 0.58774(35)$, which agrees with the best estimates 
obtained so far for polymers in the bulk, which are 
$\nu = 0.58758(7)$ (Ref.~\cite{Nickel}), $\nu = 0.5877(6)$ (Ref.~\cite{Sokal2}).

Let us consider now the ratios of eigenvalues.
For such invariant quantities 
we performed the following three-parameter fits
\be
\frac{\<q_k\>_N}{\<q_1\>_N} = L^2_{k,1} + D\, {\li{N_0\over N}\ri^\Delta}\qquad
k=1,2 \, ,
\label{rrr}
\ee
where $L^2_{k,1}$, $D$, and $\Delta$ are free parameters 
and $N_0 = 16000$ is a fixed parameter introduced 
in order to reduce the cross-correlations
between $D$ and $\Delta$ in the correction-to-scaling term.
Again, since we are neglecting higher-order corrections to scaling,
the estimates of the
universal amplitudes are affected by systematic errors.
In order to take into account this effect, we performed
fits in which only data with $N \ge N_{\rm min}$ were considered
and we studied how our estimates approach the asymptotic value as 
$N_{\rm min}$ increases.
The results of the fits are shown in Table~\ref{rapp}.
In this case the estimates from the three-parameter fits
appear to be very stable, so that we can
choose as best estimates of the universal ratios the ones for $N_{\rm min} = 1000$:
$L^2_{2,1} = 2.9272(28)$, $L^2_{3,1} = 14.6293(66) $.

Nevertheless, one can notice that, while the correction-to-scaling
exponent determined from the ratio $\<q_3\>_N/\<q_1\>_N$ is
$\Delta = 0.450(46)$, which is in agreement with the estimates
for the leading correction-to-scaling exponent for polymers in the
bulk~\cite{note}, the estimate of $\Delta$ obtained in fits
of $\< q_2\>_N/\< q_1\>_N$ is considerably smaller.
This can be due to the presence of corrections with opposite
sign and could affect our final estimate.
For this reason, we also performed fits in which the leading 
correction-to-scaling exponent was fixed at $\Delta = 0.5$, and higher-order 
corrections were neglected. The estimates obtained are shown in 
the last three rows of Table~\ref{rapp}.
Estimates for $L^2_{2,1}$ are stable for $N_{\rm min} = 4000$,
while those for $L^2_{3,1}$ are already stable for $N_{\rm min} = 1000$.
In all cases the $\chi^2$ per degree of freedom is less than 1. Therefore,
we can take as final estimates
\bea
L^2_{2,1}& =& 2.94931(99)\label{l1}\, ,\\
L^2_{3,1}& =& 14.6449(60)\label{l2}\, .
\eea
One can notice that the estimate in Eq.~\reff{l2} agrees with the one
obtained without fixing $\Delta$, while $L^2_{2,1}$ in Eq.~\reff{l1}
is not compatible with the previous estimate. 
Since we know {\em a priori} that $\Delta$ should be approximately $1/2$,
we are inclined to consider~\reff{l2} a more reliable estimate.

We performed the same kind of analysis for the asphericities, performing
fits of the form
\bea
A_3^N& =& A_3  + D\, {\li{N_0\over N}\ri^\Delta}\, ,\\
\widehat A_3^N& =&\widehat  A_3  + \widehat D\, {\li{N_0\over N}\ri^\Delta}\, .
\label{aaa}
\eea
The results are reported in Table~\ref{asf}.
Fits are not very stable and the amplitude appearing in the correction to scaling
is very small. For $N_{\rm min} = 1000$ we estimated 
$D \approx (47 \pm 10)\times 10^{-5}$ and $\widehat D \approx (43\pm 7)\times 10^{-5}$.
As before, we also performed a two-parameter
fit in which the leading correction-to-scaling exponent was fixed at $\Delta = 0.5$.
Since all estimates for $N_{\rm min} \ge 1000$ are compatible within error bars, 
our final estimates are
\bea
A_3& =& 0.44520(15)\, ,\label{a3}\\
\widehat A_3& =& 0.55949(13)\, .\label{hata3}
\eea
These estimates are in agreement with those obtained letting $\Delta$ be a free
parameter.
\begin{table}
\begin{center}
\begin{tabular}{|c||c|c|c|c|}
\hline
\hline
$N$ & $ \< q_1 \>_N \pm \delta \< q_1 \>_N $& $ \nu  \pm \delta \nu $ & $\chi^2$ & DF\\
\hline
  1000 &$  35.613 \pm 0.030 $&$  0.59000\pm  0.00009$& 153.355 &3\\
  4000 &$ 184.151 \pm 0.058 $&$   0.58949\pm  0.00010 $& 41.982&2\\
  8000 &$ 417.76 \pm 0.057 $&$ 0.58901\pm 0.00013 $&  1.294&1\\
 16000 &$  945.49 \pm 0.248 $&$  0.58867\pm 0.00033 $& 0 &0 \\
 32000 &$ 2138.31 \pm 0.818 $&&&\\
\hline
\hline
$N$ & $ \< q_2 \>_N \pm \delta \< q_2 \>_N $& $ \nu  \pm \delta \nu $ & $\chi^2$ & DF\\
\hline
  1000&$  105.71 \pm 0.10 $&$ 0.58936\pm  0.00010$&77.460 &3\\
  4000&$  545.34\pm  0.18 $&$0.58894\pm  0.00011$& 20.057 &2\\
  8000&$  1235.79 \pm 0.38 $&$0.58834\pm  0.00018$&1.348 &1\\
  16000&$  2794.68 \pm 0.78 $&$0.58799\pm  0.00035$&  0&0\\
  32000&$  6314.5 \pm 2.5 $&&&\\
\hline
\hline
\hline
$N$ & $ \< q_3 \>_N \pm \delta \< q_3 \>_N $& $ \nu  \pm \delta \nu $ & $\chi^2$ & DF\\
\hline
  1000 &$ 528.12  \pm 0.59$&$0.58881\pm  0.00011$&30.2&3\\
  4000 &$ 2714.9  \pm 1.0$&$0.58853\pm 0.00013$&8.27&2\\
  8000 &$ 6145.0  \pm 2.1$&$0.58814\pm 0.00020$&1.44&1\\
  16000 &$ 13893.5  \pm 4.3$&$0.58774\pm 0.00038$&0&0\\
  32000 &$ 31381.  \pm 13.$&&&\\
\hline
\hline
$N$ & $\<R_g^2\> \pm \delta \<R_g^2\>$& $\nu \pm \delta \nu$&$\chi^2$&DF\\
\hline
1000 & $670.51 \pm   0.36 $&$0.58918\pm 0.00009 $&44.6&3\\
4000 & $3445.4 \pm    1.3 $&$0.58857\pm 0.00013 $&4.13&2\\
8000 & $7796.6 \pm    2.4 $&$0.58830\pm 0.00020 $&0.69&1\\
16000 & $17628.2  \pm  4.8$&0.58801$\pm 0.00040 $&0&0\\
32000 & $39831  \pm  19 $  & &&\\
\hline
\end{tabular}
\end{center}
\caption{Monte Carlo results for the eigenvalues and for the trace of the
radius-of-gyration tensor $Q$ and effective exponents computed using the 
results for walks of length $\ge N_{\rm min}=N$. 
Data are fitted as in Eq.~\reff{fit1}.  ``DF'' is the number of degrees of freedom.}
\label{eigen}
\end{table}

\begin{table}
\protect\footnotesize
\begin{center}
\begin{tabular}{|c|c|c|c||c|c|c|}
\hline
\hline
$N$ & \multicolumn{3}{c}{$ \frac{\<q_2\>_N}{\<q_1\>_N}
\pm \delta \frac{\<q_2\>_N}{\<q_1\>_N} $}& \multicolumn{3}{c|}{ 
 $  \frac{\<q_3\>_N}{\<q_1\>_N}  \pm \delta \frac{\<q_3\>_N}{\<q_1\>_N} $ }\\
\hline

     1000       &   \multicolumn{3}{c||}{ $2.9683 \pm   0.0025$}       & \multicolumn{3}{c|}{$14.830 \pm   0.018$}  \\
\hline
     4000       &   \multicolumn{3}{c||}{ $2.96137 \pm   0.00073$}        & \multicolumn{3}{c|}{$14.7426 \pm   0.0052$}  \\
\hline
     8000       &   \multicolumn{3}{c||}{ $2.95804 \pm   0.00063$}        & \multicolumn{3}{c|}{$14.7101 \pm   0.0047$}  \\
\hline
    16000       &   \multicolumn{3}{c||}{ $2.95580 \pm   0.00055$}        & \multicolumn{3}{c|}{$14.6945 \pm   0.0040$}  \\
\hline
    32000       &   \multicolumn{3}{c||}{ $2.95303 \pm   0.00078$}        & \multicolumn{3}{c|}{$14.6762 \pm   0.0057$}  \\
\hline
\hline
\hline
$N_{\rm min}$ & $ L_{2,1}^2 \pm \delta L_{2,1}^2 $ & 
$ \Delta \pm \delta \Delta$ & $\chi^2/ {DF}$ & $ L_{3,1}^2 \pm \delta
L_{3,1}^2 $ &
$ \Delta \pm \delta \Delta$ & $\chi^2/ {DF}$ \\
\hline
\hline
1000 & 	$ 2.9272 \pm 0.0028$  & $ 0.150 \pm  0.017 $ & 0.167  & $  14.6293 \pm 0.0066$   &$ 0.450
\pm 0.046$ & 0.39 \\
\hline 
4000 & $2.9273 \pm 0.0032$ & $0.15 \pm 0.020 $& 0.33  & $14.6415 \pm 0.0074$ & $ 0.525 \pm 0.083 $& 
0.73\\
\hline
\hline
$N_{\rm min}$ & $ L_{2,1}^2 \pm \delta L_{2,1}^2 $ & 
$ \Delta = 0.5 $ & $\chi^2/ {DF}$ & $ L_{3,1}^2 \pm \delta
L_{3,1}^2 $ &
$ \Delta = 0.5 $ & $\chi^2/ {DF}$ \\
\hline
\hline
1000 & 	$ 2.95026 \pm   0.00082 $  & &1.36 & $14.6449 \pm   0.0060  $   &  & 0.32 \\
\hline 
4000 & $2.94931 \pm   0.00099 $&&  0.56  & $ 14.6431 \pm   0.0072 $ &  &0.37 \\
\hline 
8000 & $2.9487 \pm   0.0015 $&&  0.82  & $ 14.646 \pm   0.011  $ & &0.56 \\
\hline
\hline
\end{tabular}
\end{center}
\caption{Monte Carlo results for the ratios of eigenvalues and results
of the corresponding fits for different values of $N_{\rm min}$.
Data were fitted to the scaling form~\reff{rrr}.
In the first set of fits $\Delta$ was taken as a free parameter, while
in the second set $\Delta = 0.5$.}
\label{rapp}
\end{table}

\begin{table}
\protect\footnotesize
\begin{center}  
\begin{tabular}{|c|c|c|c||c|c|c|}
\hline
\hline
$N$ & \multicolumn{3}{c}{$ A^N_3  \pm \delta A^N_3 $ 
}& \multicolumn{3}{c|}{ 
$  \widehat A^N_3  \pm \delta \widehat A^N_3 $} \\
\hline
     1000       & \multicolumn{3}{c||}{   $0.44806 \pm   0.00033$}        & 
\multicolumn{3}{c|}{$0.56210 \pm   0.00040$ }\\
\hline
     4000       & \multicolumn{3}{c||}{   $0.44641 \pm   0.00017$}        & 
\multicolumn{3}{c|}{ $0.56104 \pm   0.00011$ }\\
\hline
     8000       & \multicolumn{3}{c||}{   $0.44623 \pm   0.00018$}        &
\multicolumn{3}{c|}{$0.56035 \pm   0.00010$}\\
\hline
    16000       & \multicolumn{3}{c||}{   $0.44604 \pm   0.00013$}        &
\multicolumn{3}{c|}{$0.56010 \pm   0.00008$ }\\
\hline
    32000       & \multicolumn{3}{c||}{   $0.44554 \pm   0.00016$}        &
\multicolumn{3}{c|}{$0.56019 \pm   0.00012$ }\\
\hline
\hline
\hline
$N_{\rm min}$ & $ A_3 \pm \delta A_3 $ & 
$ \Delta \pm \delta \Delta$ & $\chi^2/ {DF}$ & 
$ \widehat A_3 \pm \delta \widehat A_3 $ & 
$ \Delta \pm \delta \Delta$ & $\chi^2/ {DF}$ \\
\hline
\hline
1000 & 	$ 0.44520 \pm  0.00015$  & $ 0.538 \pm 0.085 $ & 1.70  & $ 0.5596 \pm 0.00011$ & $ 0.575 \pm 0.080$&  3.781 \\
\hline
\hline
$N_{\rm min}$ & $ A_3 \pm \delta A_3 $ & 
$ \Delta = 0.5 $ & $\chi^2/ {DF}$ & 
$ \widehat A_3 \pm \delta \widehat A_3 $ & 
$ \Delta = 0.5$ & $\chi^2/ {DF}$ \\
\hline
1000 &$0.44520 \pm   0.00015$&&1.12 &$0.55949 \pm   0.00013$&&2.58\\
\hline
4000 &$0.44530 \pm   0.00022$ &&1.50&$0.55945 \pm   0.00015$&&3.72\\
\hline
8000 &$0.44496 \pm   0.00035$&&1.46&$0.55989 \pm   0.00024$&&1.67\\
\hline
\end{tabular}
\end{center}
\caption{Monte Carlo results for asphericities and results
of the corresponding fits for different $N_{\rm min}$.
Data were fitted to the scaling form~\reff{aaa}.
In the first set of fits $\Delta$ was taken as a free parameter, while
in the second set $\Delta = 0.5$.}
\label{asf}
\end{table} 

\section{Comparison of the results with the case of
polymers in the bulk}
Let us compare our results with the ones for polymers in the bulk~\cite{Mazur_73,Bishop_89,
Bruns,Kremer_92,Sciutto_96, Zifferer_98,Zifferer_99,Zifferer_01,Timoshenko_02}.
An early Monte Carlo simulation~\cite{Kremer_92}
of polymers with length $N\le 220$, in which the presence of corrections
to scaling was taken into account, gave $A_3 = 0.431(2)$. 
Since then, more computing power and more efficient algorithms
became available and much longer polymers could be studied.
In Ref.~\cite{Zifferer_98}, self-avoiding walks for a single value
of $N$ ($N = 963$) were simulated using the pivot algorithm,
obtaining $A_3 = 0.43333(8)$ and $\widehat A_3 = 0.54722(9)$ for the
shape asphericities. From the data in Ref.~\cite{Zifferer_98} for the shape factors, 
we can compute the eigenvalue ratios $L^2_{3,1}= 14.1102(52)$ 
and $L^2_{2,1}= 2.9631(12)$.
Since only one value of $N$ was considered, no analysis of
the corrections to scaling could be done.
Thus, we expect the errors to be grossly underestimated.
Indeed, our results show that, for $N \approx 10^3$, the systematic error
due to the scaling corrections is much larger than the quoted statistical
error.
In Refs.~\cite{Zifferer_99,Zifferer_01} even longer walks were
considered and data were extrapolated to infinite length with
leading correction-to-scaling exponents $\Delta = 0.5$ and $\Delta = 0.46$.
The estimates for the asphericities are: $A_3 = 0.4302$ and 
$\widehat A_3 = 0.5450$ for $\Delta = 0.5$; $A_3 = 0.4303$ and
$\widehat A_3 = 0.5452$ for $\Delta = 0.46$.
From the asymptotic values for shape factors, we obtain for the eigenvalue ratios:
$L^2_{3,1}= 13.89$, $L^2_{2,1}= 2.934$ for $\Delta = 0.5$; and
$L^2_{3,1}= 13.92$ $L^2_{2,1}= 2.939$ for $\Delta = 0.46$. 
No error bars on the extrapolated values were reported in
Refs.~\cite{Zifferer_99,Zifferer_01}.
The asphericity $A_3$ has also been computed in an extensive
Monte Carlo simulation in Ref.~\cite{Sciutto_96}, 
where walks of length $N \ge 40000$ were considered.
If the data reported there are fitted assuming $\Delta = 0.5$,
one finds $A_3 = 0.43074(57)$, which agrees within error bars with
the estimate for $N = 40000$, $A_3 = 0.4309(17)$, and with
the above-reported estimates of Refs.~\cite{Zifferer_99,Zifferer_01}.
The same analysis cannot be performed for the eigenvalue ratios, since
the estimates at different $N$ are not available. Nevertheless, using Table 2  of
Ref.~\cite{Sciutto_96}, we obtain $L^2_{3,1}\approx 14.05$,
$L^2_{2,1}\approx 2.963$. Both values are slightly higher than the
corresponding ones in Refs.~\cite{Zifferer_99,Zifferer_01}.
%

Comparing our estimates~\reff{a3},~\reff{hata3}, with those reported above,
we find that, due to the presence of the surface,
$A_3$ approximately increases by $3.5\%$, while $\widehat A_3$ by
$2.6\%$. The change in the two asphericity observables is,
therefore, of the same order.

Moreover, while $L^2_{2,1}$ does not appreciably change,
$L^2_{3,1}$  increases by about $5\%$.  This shows that the change
in asphericity is mainly due to the fact that the length of 
the longest axis of the inertia ellipsoid increases with respect 
to the other two.
This is a signal of the fact that the presence of the surface gives 
the polymer directionality along the normal direction to the surface,
which we expect, on average, to coincide with the direction of the
larger axis in the inertia ellipsoid.
Nevertheless, the difference is only of a few percents. Therefore, we expect that
it can be experimentally detected only in high-precision experiments.

\end{document}